\documentclass[preprint2]{emulateapj}
\bibpunct[]{(}{)}{;}{a}{}{;}
\shorttitle{Evolution of Galactic Nuclei. I.}
\shortauthors{Matsubayashi et al.} 
\begin{document}
\title{Evolution of Galactic Nuclei. 
I. orbital evolution of IMBH}
\author{Tatsushi Matsubayashi}
\affil{Communication Science Laboratories, NTT Corporation}
\affil{2-4 Hikaridai, Seika-cho, Soraku-gun, Kyoto, 619-0237, Japan }
\email{tatsushi@cslab.kecl.ntt.co.jp}
\author{Junichiro Makino}
\affil{Department of Astronomy, University of Tokyo}
\affil{7-3-1 Hongo, Bunkyo-ku, Tokyo 113-0033, Japan}
\email{makino@astron.s.u-tokyo.ac.jp}
\and
\author{Toshikazu Ebisuzaki}
\affil{Computational Astrophysics laboratory, RIKEN}
\affil{2-1 Hirosawa, Wako-shi, Saitama, 351-0198, Japan}
\email{ebisu@riken.jp}
\begin{abstract}
Resent observations and theoretical interpretations suggest that
IMBHs (intermediate-mass black hole) are formed in the centers of
young and compact star clusters born close to the center of their
parent galaxy. Such a star cluster would sink toward the center of the
galaxy, and at the same time stars are stripped out of the cluster by
the tidal field of the parent galaxy. We investigated the orbital 
evolution of the IMBH, after its parent cluster is completely 
disrupted by the tidal field of the parent galaxy, by means of 
large-scale $N$-body simulations. We constructed a model of the central 
region of our galaxy, with an SMBH (supermassive black hole) and 
Bahcall-Wolf stellar cusp, and placed an IMBH in a circular orbit of 
radius 0.086pc. The IMBH sinks toward the SMBH through dynamical 
friction, but dynamical friction becomes ineffective when the IMBH 
reached the radius inside which the initial stellar mass is comparable 
to the IMBH mass. This is because the IMBH kicks out the stars. This 
behavior is essentially the same as the loss-cone depletion observed in
simulations of massive SMBH binaries. After the evolution through
dynamical friction stalled, the eccentricity of the orbit of the IMBH 
goes up, resulting in the strong reduction in the merging timescale 
through gravitational wave radiation. Our result indicates that the 
IMBHs formed close to the galactic center can merge with the central 
SMBH in short time. The number of merging events detectable with DECIGO 
is estimated to be around 50 per year.  Event rate for LISA would be 
similar or less, depending on the growth mode of IMBHs.
\end{abstract}
\maketitle
\cleardoublepage
\section{Introduction}
Recent observations \citep{matsumoto2001,matsushita2000} suggested 
that intermediate-mass black holes (IMBHs) exist in some starburst 
galaxies. The first such object, M82 X-1 (the brightest source in Figure 1 
of \cite{matsumoto2001}) lays 200pc off the dynamical center of M82, 
and has the estimated minimum mass of 700 $M_{\odot}$. Several 
scenarios have been proposed for the formation of this object 
\citep{ebisuzaki2001,miller2002,kawakatu2002,taniguchi2000}. 
Both Ebisuzaki et al. and Miller and Hamilton argued that IMBHs
are formed through stellar dynamical process and merging. The main 
difference between them is simply in the time at which the merging 
occurs. Ebisuzaki et al. assumed that most of merging process occurred 
while participants were main-sequence stars. On the other hand, Miller and 
Hamilton assumed that the IMBH grew through merging of smaller black holes. 

Which of the two processes actually occur depends mainly on the initial
thermal relaxation time of the cluster. \citet{simon2004} showed, 
using $N$-body simulations, that the runaway merging of massive star 
occurs if the initial relaxation time of the star cluster is less than 
4 Myrs, which is of the order of the lifetime of massive stars. The 
compact star cluster MGG-11 \citep{mccrady2003}, which coincides with 
the location of M82 X-1, has the estimated relaxation time of 3 Myrs. 
On the other hand, the relaxation time of MGG9, which is more massive 
than MGG11, is significantly longer. This difference is consistent with 
the existence and nonexistence of IMBHs in MGG-11 and 9, respectively. 
In the case of the scenario by Miller and Hamilton, the growth timescale 
would be much longer, and it is hard to explain why any IMBH can exist 
in a young cluster like MGG-11. 

Compared to the relaxation time of typical globular clusters, which is
$10^{8\sim 9}$ yrs, the relaxation time of a few Myrs might sounds
extremely short. However, for young clusters, such short relaxation
time is not unusual. For example, 
Arches and Quintuplet clusters, whose estimated ages are around 1-5 Myrs
years, have the estimated relaxation time of 12M years 
\citep{simon2002}. Star clusters R136 in LMC and Westerlund 1 in our 
galaxy are other examples of such young compact clusters.

If the formation of IMBHs is not a rare event in young and compact
clusters, two questions naturally arise. What is the final fate of
the IMBH and its parent cluster, and whether or not it is related to
the growth of the central black hole of the parent galaxy. The star
cluster itself evolves through internal thermal relaxation, tidal
truncation, and dynamical friction, much in the same way as globular
clusters evolve. The main difference is again that the timescale is much
shorter. For example, if there is a cluster with mass $10^5M_{\odot}$
at the distance of 30 pc from the center of our Galaxy, the dynamical
friction timescale would be 
\begin{eqnarray}
t_{\rm fric} & \simeq &
\frac{2.38 \times 10^8}{\ln \Lambda} 
\left( \frac{r}{30 {\rm pc}} \right)^2
\nonumber \\
&&
\left( \frac{\sigma}{100 {\rm km~s^{-1}}} \right)
\left( \frac{10^5M_\odot}{ M_c} \right) yr~,
\label{eq:t-fric}
\end{eqnarray}
where $\sigma$ is a velocity dispersion and $M_c$ is the mass of the cluster.

As the cluster sinks toward the center of the parent galaxy, the tidal
field of the galaxy becomes stronger. As a result, the cluster loses mass, and
eventually becomes completely disrupted, leaving the IMBH orbiting around the
central SMBH. IRS13E \citep{maillard2004} looks like such a remnant
cluster, composed of a single IMBH and several stars still bound to it
\citep{simon2005}
IRS13E is located at 4 arcsec ($\sim$0.16pc) from Sgr A$^{\ast}$. 
It appears as a cluster 
of seven individual stars within a projected diameter of $\sim$ 
0.5 arcsec (0.02 pc). All these sources have a common westward 
proper motion, indicating that they are bound (at least six of them). 
The types of stars imply that it is a young star cluster with the age
of a few Myrs. To keep these stars bound, the total gravitational
mass of IRS13E must exceed $1300 M_\odot$, about one order of magnitude
larger than the estimated total mass of visible stars. One natural
interpretation is that IRS13E is a remaining core of much more
massive star cluster, with an IMBH of $\sim 1300 M_\odot$.

In this paper, we consider the orbital evolution of an IMBH after its
parent cluster is completely disrupted. The main question is the
merging timescale of the IMBH and the central SMBH. In the case of a massive 
BH binary, which 
forms when two galaxies each with a central massive BH merge,
the merging timescale has been the area of active research since the
pioneering work by \citet{BBR1980}. In this case, the result of
recent large-scale $N$-body simulations \citep{makino2004} suggests 
that the merging timescale is much longer than the Hubble time.
They demonstrated that the evolution timescale of the BH binary is
proportional to the relaxation time of the parent galaxy, as suggested
by \citet{BBR1980}. This conclusion is different from the results of
previous simulations \citep{quinlan1997,makino1997,milos2001,milos2003,
chatterjee2003}, but the difference is mostly due to the limitation 
in the number of particles in previous simulations. \citet{szell2005} 
and \citet{berczik2005} obtained similar results as \citet{makino2004}, 
albeit with somewhat smaller number of particles.

However, we cannot directly apply these result to the case of an
SMBH-IMBH binary, because of its very large mass ratio. In the case of
SMBH-SMBH binary, the binary need to interact with the field stars
with the total mass comparable to that of the total mass of the binary
to change its internal orbital parameters significantly. Thus, the
loss-cone depletion occurs when the binary ejected out the mass
comparable to its mass, and the structure outside the loss cone region
is self-gravitating. In the case of SMBH-IMBH binary, the IMBH orbit can
evolve by interacting with the mass comparable to the IMBH mass which
is several orders of magnitude smaller than the SMBH mass. Thus, loss
cone depletion can occur when the IMBH ejected out the central stellar
mass comparable to the IMBH mass, and the gravitational potential of
this region, or actually of the region much further out, is dominated
by the potential of the central BH. Thus, unlike the case of SMBH-SMBH
binary, we need to study the evolution of a highly unequal mass
binary, in the distribution of stars which are all bound to the
primary component of the binary. 

The organization of the paper is as follows. In \S 2, we describe the 
numerical method and the initial conditions we used. In \S 3 and \S 4, 
we describe the result of the simulations. Summary and discussion are 
presented in \S 5.

\section{Initial models and numerical methods}
\label{methods}

\subsection{Initial Models}
\label{sec:initialconditions}
The goal of this paper is to study the orbital evolution of an IMBH in
the stellar distribution where the gravitational potential of the
central SMBH dominates. As the background stellar distribution, we
adopt the standard $\rho \propto r^{-7/4}$ cusp \citep{bahcall1976}. 
One practical problem with this distribution is that the total mass 
would be infinity if the density profile is given by this single power 
law. In addition, such a model is physically unacceptable, since the 
basic assumption for the Bahcall-Wolf cusp is that the gravitational 
potential is dominated by that of the central BH.

In order to construct a model with central density slope of $-7/4$ and
finite mass, we use Tremaine's $\eta$-model with central BH
\citep{tremaine1994}, with one modification. The original 
$\eta$-model has the outer slope of $-4$. We constructed the
model with outer slope of $-5$, just to make the mass in the outer
region smaller.

The density distribution of our modified model is given by 
\begin{eqnarray}
\rho_{\eta}(r) = \frac{\eta}{4\pi}
\frac{M_{\eta}r_{0}^{2}}{r^{3-\eta}(r_{0}^{2}+r^{2})^{\eta/2 +1}}~,
~0<\eta \le 3 ,
\label{eq:rho}
\end{eqnarray}
where $M_{\eta}$ is the total mass of field stars and $r_{0}$ is the 
scale length. The model with $\eta = 5/4$ correspond to profile 
$\rho \propto r^{-7/4}$, and we use it in this paper. 

In order to construct an $N$-body model in dynamical equilibrium,
we need to construct the distribution function (DF). We could obtain
DF at least numerically by solving the Abel integral
equation. However, since what we need is a model with an inner power-law
cusp and some outer cutoff, it would be an overkill to obtain a
distribution function which exactly satisfies equation (\ref{eq:rho}).
So we approximate DF by the following formula.
\begin{eqnarray}
f(\epsilon) = f_{0}(\epsilon)^{7/2}
\left(\epsilon_{0}^s+\epsilon^{s}\right)^{-\frac{\eta +2}{s}}~,
\label{eq:f-e}
\end{eqnarray}
where $\epsilon \equiv -E$, 
\begin{eqnarray}
\epsilon_0=\left(\frac{f_{1}}{f_{0}}\right)^{-1/(\eta +2)}~,
\label{eq:e-0}
\end{eqnarray}
\begin{eqnarray}
f_{0}=\frac{\eta M_{\eta}\Gamma (4-\eta)}
{2^{7/2}\pi^{5/2}M_{S}^{3-\eta}\Gamma (5/2 -\eta)}, 
\label{eq:f-0}
\end{eqnarray}
and 
\begin{eqnarray}
f_{1}=\frac{8\sqrt{2}}{7\pi^{3}}{\eta M_{\eta}}~.
\label{eq:f-1}
\end{eqnarray}

DF of this form gives a correct asymptotic behavior for both $\epsilon
\rightarrow \infty$ and $\epsilon \rightarrow 0$. This formula has one
adjustable parameter $s$. We chose $s=5$ after some numerical tests. 

We generated the initial $N$-body model in the following two
steps. First, we generate positions of $N$ particles so that the
density distribution obeys equation (\ref{eq:rho}). Then we assign the
velocity to each particle, so that the velocity distribution at the
point of particle is consistent with the DF given in equation
(\ref{eq:f-e}). Since the DF we used is an approximate solution, the
constructed $N$-body model is not in an exact dynamical
equilibrium. However, as we will see in \S \ref{sec:stability},
our initial model is in pretty good dynamical equilibrium.

We chose the system of units in which the total mass
of the system and gravitational constant are both unity and the total
binding energy of the system is $-1/4$. We set the mass of SMBH to be
$M_S=0.5$, that of the IMBH $M_I = 5\times 10^{-4}$. The mass of the
stellar distribution is therefore $0.4995$. The stellar mass 
inside the radius $r$ is 
\begin{eqnarray}
M_{\eta}(r) = M_\eta\frac{r^{\eta}}{(r_{0}^{2}+r^{2})^{\eta/2}}. 
\label{eq:Mr}
\end{eqnarray}
For the above choice of the system of units, $r_0 \simeq 2.39$. When
constructing the initial stellar distribution, we exclude the stars
with periastron distance less than $10^{-3}$, to avoid numerical
problem.

\begin{figure}[htpb]
\plotone{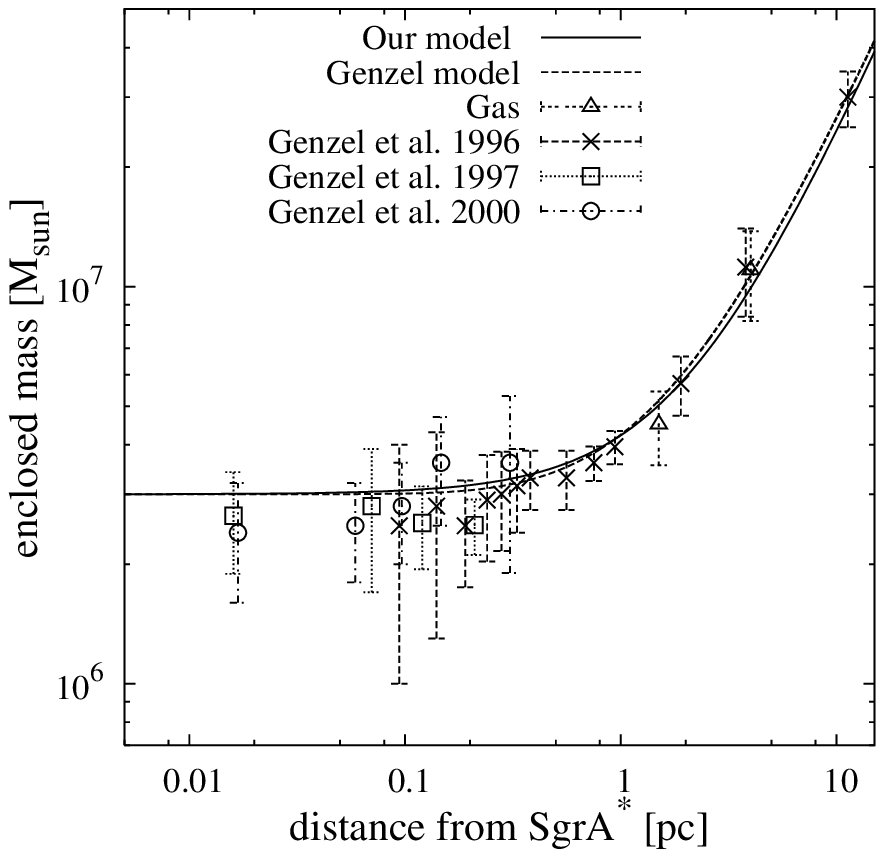}
\caption{Cumulative mass distribution of Our Galaxy 
obtained from stellar and gas dynamics and that of 
our and Genzel 2000 models. Triangles denote mass 
estimates from the gas dynamics \citep{guesten1987}. 
Crosses, circles and squares denote the isotropic 
mass modeling of \citet{genzel1996,genzel1997,genzel2000}. 
Dashed curve is the mass model by \citet{genzel2000}, 
whose stellar distribution is 
$\rho(r)=\rho_0 [1+(r/r_0)^2]^{-\alpha/2}$, of 
$\rho_0=3.5\times 10^6 M_\odot$, $r_0=0.17$pc and 
$\alpha =1.8$. Solid curve is our model, drawn 
using single power-law mass distribution with 
$\rho \propto r^{-7/4}$.
\label{fig:genzel}}
\end{figure}

To convert the timescale obtained in our simulations to the physical
timescale, we need to define conversions between our system of units
and real physical units. We use unit mass $M=6.0\times 10^6 M_\odot$,
unit length $R=0.86$pc, resulting in the unit time of $T=4.6\times
10^3 $yrs. Mass and length units are chosen so that the stellar mass
inside the radius 0.7pc (assuming that the single power law with
$-7/4$ slope continues to that radius) is $7.3\times 10^5
M_\odot$ \citep{genzel2000}. Figure \ref{fig:genzel} shows the mass
distribution for our model (assuming single power-law density),
the model by \citet{genzel2000}, and observational data. Note that our
model is within the observational error bars and practically
indistinguishable with the model by \citet{genzel2000}.

The IMBH particle is placed in a circular orbit at the radius 0.1 
(0.086pc in physical units). Its mass is $3000M_\odot$.

We performed four runs with different number of field particles (models
A1 to A4 of table 1). The mass of a field particle in model A4 is
30 $M_\odot$, which is not too far from the mass of real stars
(whatever they are) in this region.

In table 1, models B1 to B4 were prepared to study the evolution of
the IMBH after it approaches to less than 0.01 pc to the SMBH. For these
models, we used $r_0$ much smaller than that is used for model Ax, to
reduce the total number of stars. We used $r_0=0.1$ and placed the IMBH at
$r=0.01$, so that the stellar distribution at the initial position of
IMBH is still the power law with slope $-7/4$. The total stellar mass
is $8.9\times 10^{-3}$, or about 20 times the IMBH mass. Results of
Bx models will be discussed in \S 4. Mass of a field particle in model
B4 is 3 $M_\odot$, which is of the same order with that of real
stars in this region. Thus, two-body relaxation effect in this model
is essentially the same as what would occur in the real galactic center.

We made models C1 and C2 to test the effect of softening parameters
(see in \S \ref{sec:hardware}) and D1 and D2 to test the validity of 
the initial model and relaxation effect (see \S \ref{sec:stability}).

\subsection{Hardware and Numerical Integration Method}
\label{sec:hardware}
For all calculations, we used simple direct-summation algorithm and 
fourth-order Hermite scheme integrator with individual (block) 
time step \citep{makino1992}.

For the calculation of gravitational forces from field particles (to
both the field particle and black holes), we used the GRAPE-6 (and
-6A) of Tokyo University, a special purpose computer for $N-$body
simulation \citep{makino2003,fukushige2005}. The calculation of
the forces from black holes was done on the host computer to maintain
sufficient accuracy. We did not include any relativistic effects and
treated BH particle as massive Newtonian particle.

In our calculation, we did not use regularization technique to keep
the calculation code simple. Thus, we need to apply softening
parameters for gravitational interactions between particles. 
We use four different softening lengths, for BH-BH, SMBH-star,
IMBH-star, and star-star interactions. 

For SMBH-IMBH interaction, we apply zero softening. They cannot
easily come close enough to each other for the numerical difficulty to
occur. If such close encounter occurs, the timescale of orbital
evolution through gravitational wave radiation becomes short enough so
that our pure Newtonian treatment is not really valid. 

The softening for BH-star interaction should be determined with some
care, since close encounters do occur and too large softening can
affect the orbital evolution of the IMBH. For the softening of SMBH-star
interaction, we used $\epsilon_{S\mbox{-}s} = 10^{-4}$ and $10^{-6}$ in Ax
and Bx runs, respectively. These values are chosen so that the effect
of softening is small enough for stars at the radius comparable to
that of the IMBH.

For IMBH-star interaction, we chose the softening so that it is
smaller than 90-degree turnaround distance by at least two orders of
magnitude at the initial condition. As the IMBH approaches to SMBH, 
the velocity dispersion becomes larger, the difference between 
the softening length and the 90-degree turnaround distance becomes 
smaller, but it was always kept significantly larger than 1.

\begin{table}
\begin{center}
\caption{
List of parameters for the simulations in this paper.\label{table1}}
\begin{tabular}{crccccccc}
\tableline
\tableline
 Run & 
$N_{fs}$     \tablenotemark{a} & 
${M_I / m_\ast}$  \tablenotemark{b} & 
$r_{\rm I}$  \tablenotemark{d} & 
$\epsilon_{S\mbox{-}s}$ \tablenotemark{e} & 
$\epsilon_{I\mbox{-}s}$ \tablenotemark{f} & 
$\epsilon_{s\mbox{-}s}$ \tablenotemark{g} \\
\tableline
A1 &  9990 &  10.0 & 0.1 &$10^{-4}$&$10^{-6}$&$10^{-4}$ \\
A2 & 19980 &  20.0 & 0.1 &$10^{-4}$&$10^{-6}$&$10^{-4}$ \\
A3 & 39960 &  40.0 & 0.1 &$10^{-4}$&$10^{-6}$&$10^{-4}$ \\
A4 & 99900 & 100.0 & 0.1 &$10^{-4}$&$10^{-6}$&$10^{-4}$ \\
B1 &  1795 & 100.0 & 0.01 &$10^{-6}$&$10^{-8}$&$10^{-4}$ \\
B2 &  3589 & 200.0 & 0.01 &$10^{-6}$&$10^{-8}$&$10^{-4}$ \\
B3 &  7177 & 400.0 & 0.01 &$10^{-6}$&$10^{-8}$&$10^{-4}$ \\
B4 & 17942 &1000.0 & 0.01 &$10^{-6}$&$10^{-8}$&$10^{-4}$ \\ 
\\
C1 & 99900 & 100.0 & 0.1 &$10^{-4}$&$10^{-6}$&$10^{-3}$ \\
C2 & 99900 & 100.0 & 0.1 &$10^{-4}$&$10^{-6}$&$10^{-5}$ \\ 
\\
D1\tablenotemark{h} & 10000 &--&--&$10^{-4}$&--&$10^{-4}$ \\
D2\tablenotemark{h} & 100000 &--&--&$10^{-4}$&--&$10^{-4}$ \\
\tableline
\tableline
\tablenotetext{a}{Number of field particles.}
\tablenotetext{b}{Rasio of mass of IMBH and field star.}
\tablenotetext{d}{Initial separation of IMBH.}
\tablenotetext{e}{Softening of SMBH-star interaction.}
\tablenotetext{f}{Softening of IMBH-star interaction.}
\tablenotetext{g}{Softening of star-star interaction.}
\tablenotetext{h}{Runs D1 and D2 are SMBH and Field stars only.}
\end{tabular}
\end{center}
\end{table}

The criterion of the softening for star-star interaction is more
complicated. Since the "stars" in our model is still significantly
heavier than real stars (assuming we know what the mass of real stars
in this region), it is desirable to use large softening to reduce
the relaxation effect. On the other hand, the softening should be
small enough not to affect the distribution of stars. For most of runs
we used the softening length of $10^{-4}$. As the test calculations,
we performed two runs with $\epsilon_{s\mbox{-}s} = 10^{-3}$ and $10^{-5}$, 
(runs C1 and C2). As will be discussed in \S \ref{sec:stability}, 
these runs gave essentially the same results are the standard run 
(run A4). So we used $\epsilon_{s\mbox{-}s} = 10^{-4}$ for all other
runs. Table \ref{table1} also gives the softening parameters used. 

The largest calculation (model B4) took about five weeks on a
single-host, single processor-board GRAPE-6 system with a peak speed 
of 1 T-flops. For all calculations, the total energy is conserved to
better than 0.5\% for all Ax, Cx and Dx runs. For models Bx the total 
energy conservation is shown in Figure \ref{fig:2nd-energy}.

\subsection{Stability and relaxation effect}
\label{sec:stability}

\begin{figure}[htpb]
\plotone{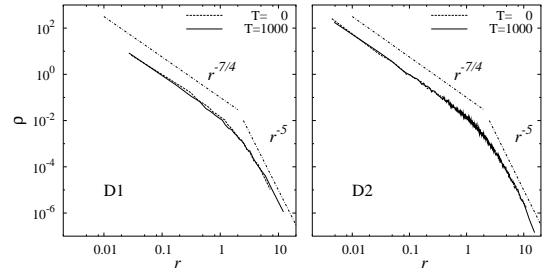}
\caption{Radial density profiles at times 
$T=0$ and $1000$ for runs D1 and D2. 
\label{fig:rho}}
\end{figure}

\begin{figure}[htpb]
\plotone{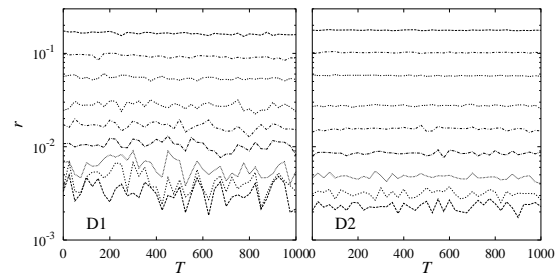}
\caption{Lagrangian radii around SMBH for runs 
D1 and D2. From bottom to top, the radii enclose 
$5\times10^{-5},$
$10^{-4}, 2\times10^{-4}, 5\times10^{-4},$
$10^{-3}, 2\times10^{-3}, 5\times10^{-3},$
0.01 and 0.02 of the unit mass. 
\label{fig:dense}}
\end{figure}
As described in \S 2.1 our initial model is not in exact dynamical
equilibrium. In addition, the system would evolve through two-body
relaxation even if the initial model is in exact dynamical
equilibrium. To see these effects, we performed two test calculations
(models D1 and D2), where we let the system without the IMBH to evolve 
for 1000 time units. Figures \ref{fig:rho} and \ref{fig:dense} show the
result.

We can see that the density profiles are practically unchanged, and
that Lagrangian radii do not show any systematic evolution. 

\subsection{Effect of star-star softening}
\label{sec:softening}

As we discussed in \S 2.2, the choice of softening for star-star
interaction might have some unpredictable effect on the evolution of
the orbit of the IMBH. To see if there is any such effect, we performed
two runs, models C1 and C2, which started from the same initial condition as
model A4 but with different values of $\epsilon_{s\mbox{-}s}$.

\epsscale{0.75}
\begin{figure}[htpb]
\plotone{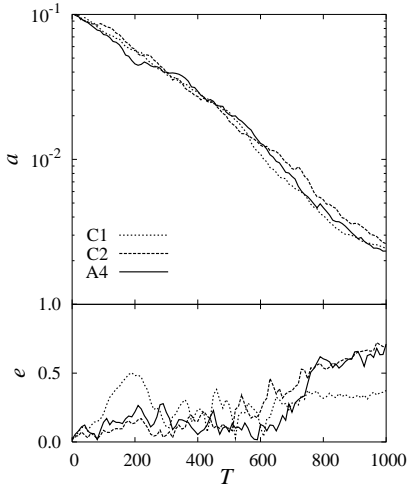}
\caption{Evolution of the semi-major axis and 
the eccentricities of the black hole binary. 
Dotted, dashed, and solid curves denote 
$\epsilon_{s\mbox{-}s} = 10^{-3}$, 
$10^{-5}$ and $10^{-4}$, respectively.
\label{fig:diff-epsiron}}
\end{figure}
\epsscale{1.0}
The result is shown in figure \ref{fig:diff-epsiron}. There is no
systematic difference among these three runs. So we can conclude that
the choice of $\epsilon_{s\mbox{-}s}$ has no significant effect on the
result.

\section{Results}
\label{sec:results-1st}

\subsection{Hardening Rate}
\label{sec:hardening-rate}

\begin{figure}[htpb]
\plotone{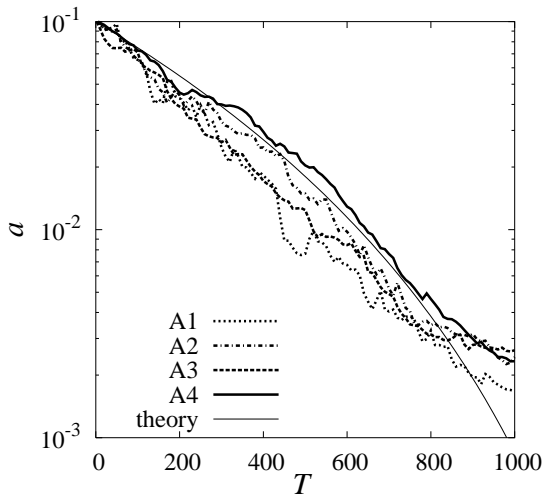}
\caption{Evolution of the semi-major axis of the IMBH.
Dotted, dash-dotted, dashed, and solid 
curves are the results of models A1-4. 
Thin curve shows the theoretical curve calculated 
equation (\ref{eq:theoretical_a}). 
\label{fig:1st-a}}
\end{figure}

Figure \ref{fig:1st-a} shows the time evolution of the semi-major axis
of the IMBH, or the SMBH-IMBH binary. Here and hereafter, we refer to
orbital elements and other quantities of IMBH-SMBH binary as those of
the IMBH, for simplicity. The semi-major axis of the IMBH is given by
\begin{eqnarray}
a = - \frac{G M_S M_I}{ 2 E_b }~,
\label{eq:a}
\end{eqnarray}
where $E_b$ is the binding energy of the IMBH
\begin{eqnarray}
E_{b} = \frac{1}{2}\mu V_{b}^{2}-\frac{G M_S M_I}{r}.
\label{eq:E-b}
\end{eqnarray}
Here, $V_{b}$ is the relative velocity of the two BHs and
$\mu$ is the reduced mass defined as
\begin{equation}
\mu \equiv M_{\rm S}M_{\rm I}/(M_{\rm S}+M_{\rm I}).
\end{equation}

From figure \ref{fig:1st-a}, we can see that the orbital evolution of
the IMBH is practically independent of the number of stars in the parent
galaxy. In all models, the IMBH is much more massive than the field
stars. So this result is not surprising.

The thin solid curve of in figure \ref{fig:1st-a} is the theoretical
prediction for the evolution of the IMBH orbit, obtained using standard
dynamical friction formula \citep{BT} 
\begin{eqnarray}
\frac{d V_b}{dt} = - 
\frac{4 \pi \ln \Lambda G^2 \rho M_I}{V_b^2} G(X).
\label{eq:hardening-rate1}
\end{eqnarray}
where 
\begin{equation}
G(X) = {\rm erf} (X) - \frac{2X}{\sqrt{\pi}} \exp (-X^{2}),
\end{equation}
and
\begin{equation}
X \equiv V_b/(\sqrt{2}\sigma).
\end{equation}
Here $\sigma$ is the velocity dispersion and we used the circular 
velocity $V_b = \sqrt{G M_S / a}$ as the velocity of the IMBH. 
We used $\ln \Lambda = 8$ here. For the calculation of dynamical
friction in inhomogeneous background distribution, it has been
suggested that taking the outer cutoff of Coulombs logarithm to the
distance of the object from the center of the parent stellar system
gives good estimate \citep{hashimoto2003}. The lower cutoff is
90-degree turnaround distance, which is given by
\begin{equation}
r_c \sim a \frac{M_I}{M_S} = 10^{-3} a.
\end{equation}
Thus, we have $\log \Lambda \sim 7$ independent of the location of
the IMBH. Using equation (\ref{eq:rho}), 
equation (\ref{eq:hardening-rate1}) can be rewritten as follows 
\begin{eqnarray}
\frac{da}{dt} &\simeq& 
\frac{10.7 \ln \Lambda G^{1/2}\rho_0 M_I a^{3/4}}{{M_S}^{3/2}} ~,
\label{eq:hardening-rate2}
\end{eqnarray}
where $\rho_0$ is the stellar mass density at $r=1$ and we assume 
the circular motion so that we used $a=r$.

We chose $X=1$ here. This differential equation has the analytic
solution given by
\begin{equation}
a = a_0 \left(\frac{T_0-T}{T_0}\right)^4~,
\label{eq:theoretical_a} 
\end{equation}
where 
\begin{equation}
T_0 = \frac{0.37~a_0^{1/4}~M_S^{3/2}}{\ln \Lambda ~G^{1/2}~\rho_0 M_I}~. 
\end{equation}
We can see that the agreement between the theoretical prediction and
numerical result is pretty good, at least for the early period ($T<600$).
In the later phase, it seems the numerical results show the slowing
down of the evolution. 

\begin{figure}[htpb]
\plotone{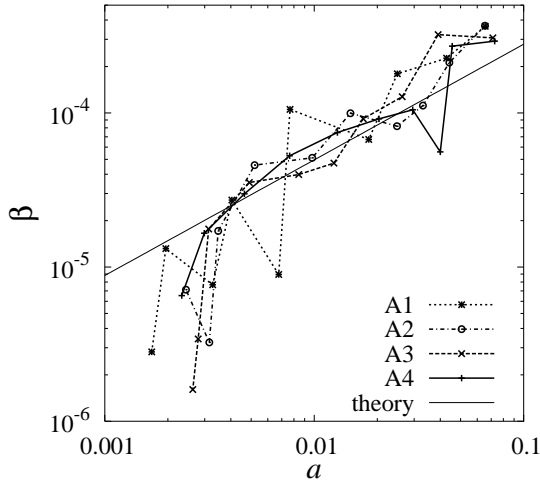}
\caption{Hardening rate $\beta$ in equation 
(\ref{eq:beta}), plotted as a function of 
semi-major axis $a$ at time $(t_0+\Delta t/2)$, 
for runs A1-4. Thin curve shows the theoretical prediction in 
equation (\ref{eq:hardening-rate2}).
\label{fig:hardening-rate}}
\end{figure}

To see the difference between the theoretical prediction and numerical 
results more clearly, we calculated the hardening rate $\beta$, 
defined as 
\begin{eqnarray}
\beta = -\frac{\Delta a}{\Delta t}. 
\label{eq:beta}
\end{eqnarray}
Here $\Delta (1/a) = |a_{1}-a_{0}|$, where $a_{1}$ and $a_{0}$ are
the semi-major axis of the IMBH at times $t=t_{0}$ and $t_{0}+\Delta t$,
respectively. We use $\Delta t = 100$ for all values of
$t_{0}$. Figure \ref{fig:hardening-rate} shows the result.
The agreement between the theoretical prediction and numerical results
is fairly good for $a>0.01$. For $a<0.01$, numerical result gives the
hardening rate smaller than the theoretical prediction, and the
difference becomes larger as $a$ becomes smaller.

\begin{figure}[htpb]
\plotone{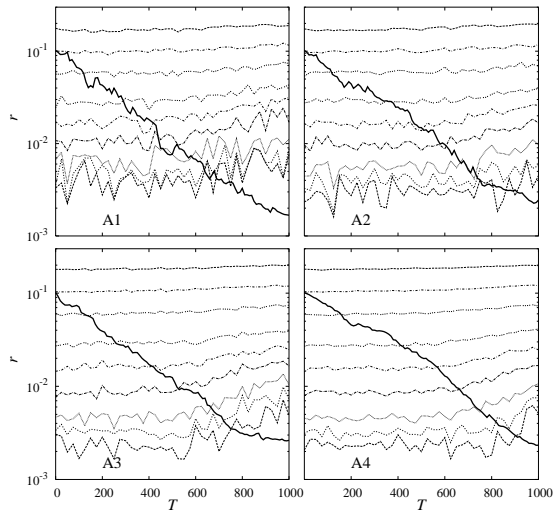}
\caption{Time evolution of Lagrangian radii of 
field stars. From bottom to top, the curves show 
the radii containing the mass $5\times10^{-5}$ 
($10^{-4}$ of the stellar mass), 
$10^{-4}, 2\times10^{-4}, 5\times10^{-4},$
$10^{-3}, 2\times10^{-3}, 5\times10^{-3},$
0.01 and 0.02. Four panels show the result 
of runs A1 through A4 (model names are shown in panels). 
Thick solid curves show the semi-major axis of the IMBH.
\label{fig:1st-dense}}
\end{figure}

A natural explanation of this slowing down is the loss-cone depletion,
similar to what happens in the case of massive BH binaries. 
Figures \ref{fig:1st-dense} show the Lagrangian radii of field
stars. We used the position of SMBH as the coordinate center.
We can see that as the IMBH sink toward the center, the Lagrangian radius
corresponding to the position of the IMBH starts to expand. For example,
in the case of model A4, the radius enclosing the mass of $2\times
10^{-3}$ (fourth curve from the top) starts to expand at around
$t=400$, which is the time the IMBH semi-major axis crosses that radius.
Radii enclosing smaller masses show similar tendency, though the
expansion is faster for radii with smaller mass.

Since the expansion of a Lagrangian radius starts when the IMBH reaches
that radius, we can conclude that this expansion is due to the back
reaction of dynamical friction to the IMBH. Thus, when the stellar mass
inside the IMBH semi-major axis becomes comparable to the IMBH mass,
the effect of back reaction becomes significant. Stellar mass inside
$r=0.01$ is about 0.1\% of the total stellar mass, which is about the
same as the IMBH mass. Thus, when the IMBH reaches the radius 0.01, the
effect of the IMBH to the stellar distribution becomes significant, and
number density of field stars is reduced. This is the reason why the
hardening rate becomes smaller when $a$ reaches 0.01.

\subsection{Change of the distribution of field stars}

\begin{figure}[htpb]
\plotone{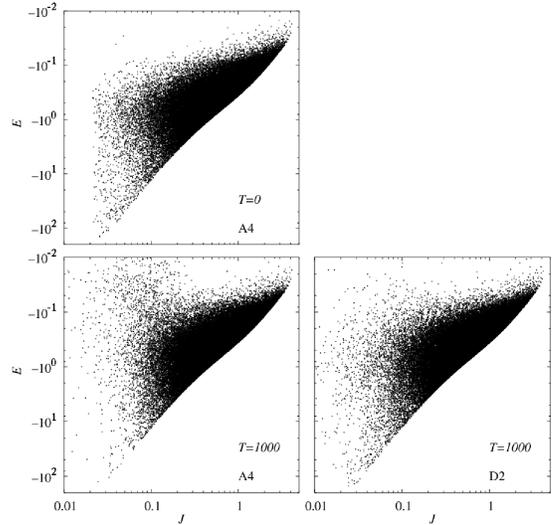}
\caption{Distribution of the field stars in 
the $(J, E)$ plane at time $T=0$ for run A4 
and those at $T=1000$ for runs A4 and D2. 
\label{fig:1st-ej}}
\end{figure}

Figures \ref{fig:1st-ej} show the distribution of the field stars in the 
$(J, E)$ plane, for run A4 and D2 at $T=0$ and 1000. Here, $E$ and $J$
are the specific binding energy and specific total angular momentum of
field stars. We take the center of mass of SMBH-IMBH binary as the origin 
tocalculate $E$ and $J$.

Note that we excluded field particles with the periastron distance to
SMBH less than $10^{-3}$ when we constructed the initial condition. 
The left-hand-side cutoff in the distribution (at around $J=0.02$) is 
due to this exclusion.

When we compare the panels, it is clear that field particles with small 
$J$ ($J< 0.03$) and large negative $E$ ($E < -10$) are depleted in model 
A4, while no such tendency is visible for model D2 (without IMBH). 
These particles are kicked out to high-energy, low-angular-momentum 
orbits by interaction with the IMBH.

There are many particles in the area $J< 0.1$ and $E> -0.2$ in $T=1000$ 
panel for run A4, while these are not in model D2. 
It is clear that these particles were kicked out by interaction with
the IMBH from the small $J$ and large negative $E$ region. 

\begin{figure}[htpb]
\plotone{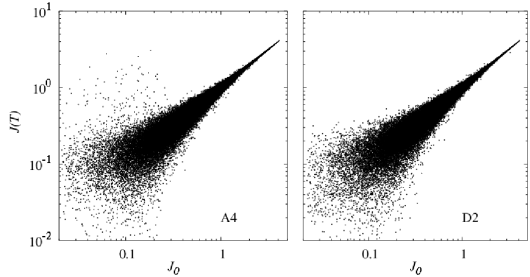}
\caption{Distribution of the field stars in 
the ($J_{0}, J(T)$) planes for runs A4 and D2. 
$J_0$ is the initial specific total angular 
momentum and $J(T)$ is that at the time $T=1000$.
\label{fig:1st-jj}}
\end{figure}

Figures \ref{fig:1st-jj} show the initial and final ($T=1000$) total 
angular momentum of stars for runs A4 and D2. In the case of run A2, 
we can see that a fair number of particles which initially have small 
$J$ ($J<0.2$) got much larger $J$. 
It indicates that field particles are kicked out by the IMBH. 
The dispersion in the case of run D2 is purely due to the
two-body relaxation.

\begin{figure}[htpb]
\plottwo{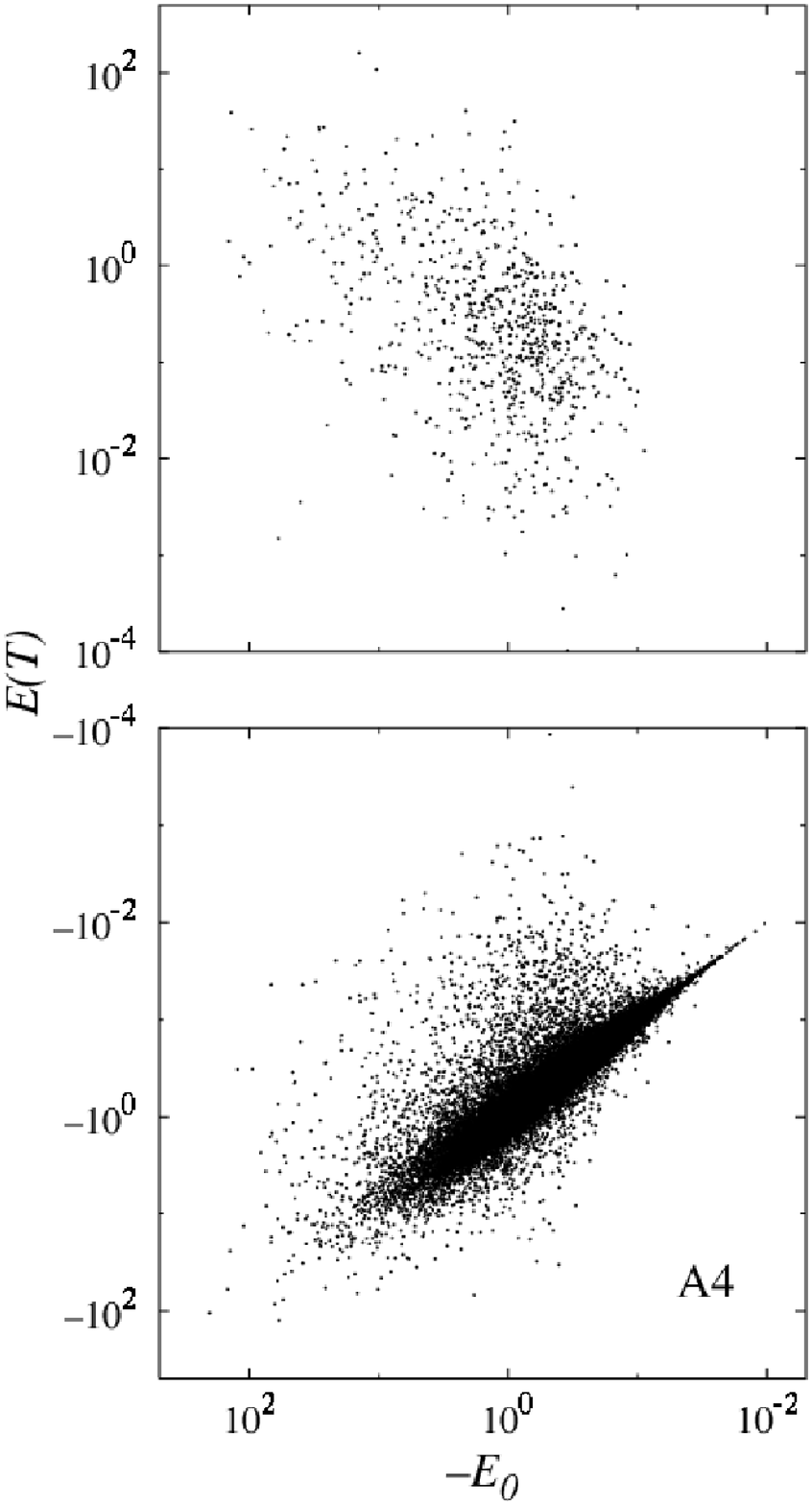}{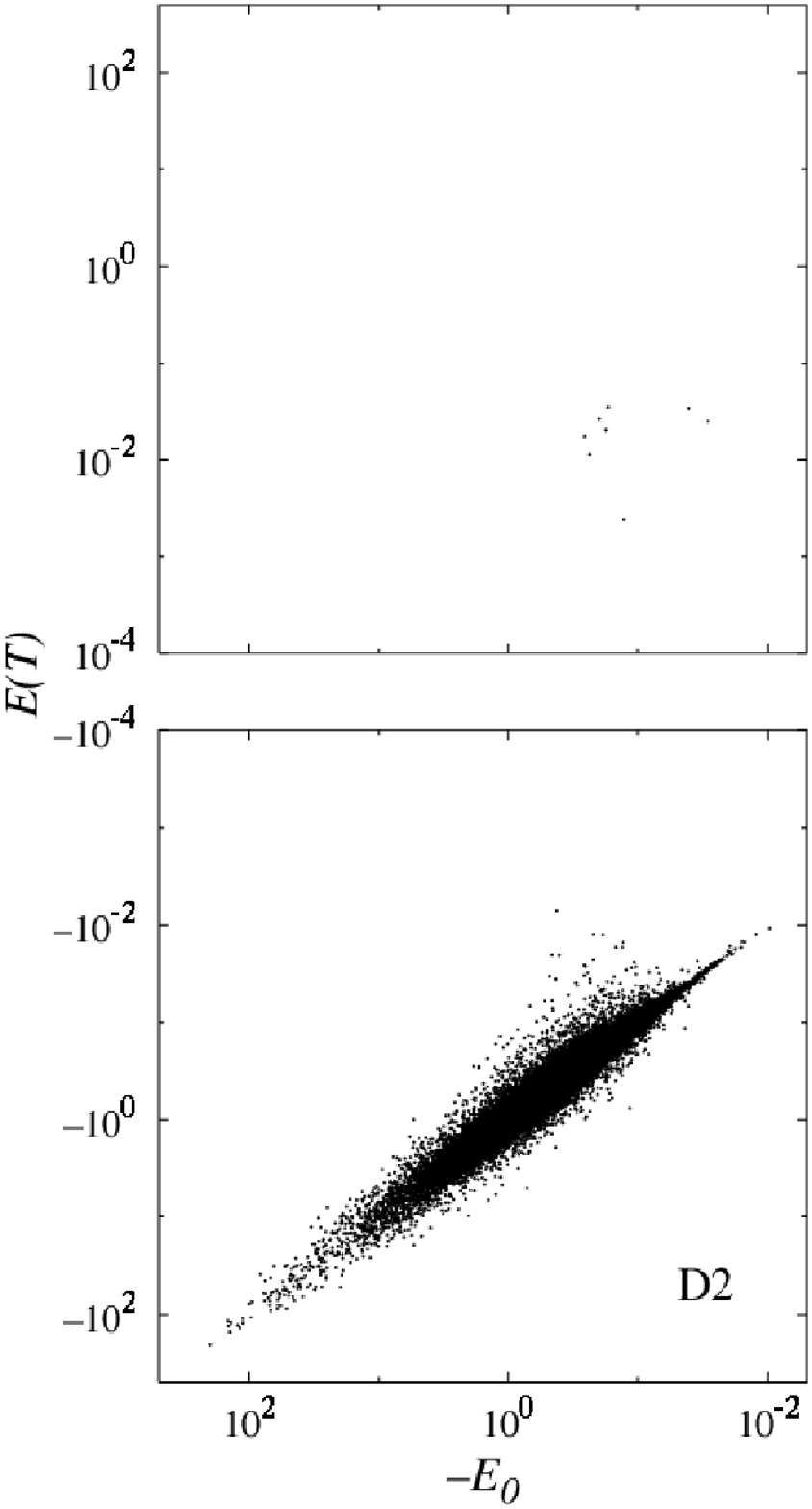}
\caption{Distribution of the field stars in 
the ($E_{0}, E(T)$) planes for runs A4 and D2. 
$E_0$ is the initial specific energy and $E(T)$ 
is that at the time $T=1000$. Top and bottom 
figure shows the distribution having a positive 
and a negative energy, respectively.
\label{fig:1st-ee}}
\end{figure}

Figures \ref{fig:1st-ee} show the initial and final binding energies
of particles for runs A4 and D2. Here, we can see the scatter is much
larger for run A4.

\subsection{Evolution of eccentricity}
\label{sec:evolution-e}

\vspace*{10mm}
\epsscale{0.5}
\begin{figure}[htpb]
\plotone{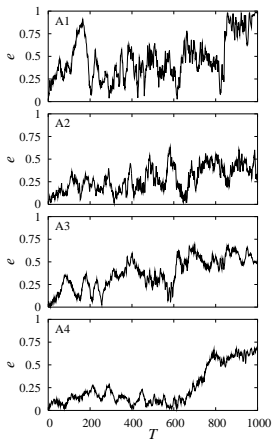}
\caption{Evolution of the eccentricity of the 
IMBH for runs Ax.
\label{fig:1st-e}}
\end{figure}
\epsscale{1.0}

Figure \ref{fig:1st-e} shows the time evolution of the 
IMBH-SMBH binary eccentricity defined as 
\begin{eqnarray}
e = \sqrt{1 - \frac{L^{2}}{(M_S+M_I)a}}~,
\label{eq:ecc}
\end{eqnarray}
where $L$ is the angular momentum of the binary.

From these results, it is not clear if there is any systematic change
in $e$ or whether or not the evolution depends on $N$. The
fluctuation in $e$ is bigger for small-$N$ calculation (A1 and A2 
compared to A3 or A4). On the other hand, in run A4 $e$ seems to show 
systematic increase after $T=600$. As we can see from figure 
\ref{fig:1st-a}, it may be related to the slowing down of the evolution 
of the semi-major axis, which is caused by the loss-cone depletion.

In the next section, we will investigate this evolution of the IMBH 
orbit after the depletion of the loss cone.

\section{The evolution of IMBH orbit after the loss-cone depletion}
\label{sec:results-2nd}
In the previous section, we have seen that the IMBH sink toward the SMBH
through dynamical friction, and its orbital evolution slows down when
the IMBH reaches the radius total stellar mass inside which is comparable
to the IMBH mass. In other words, it occurs after the IMBH ejected 
out the neighboring stars. In this section, 
we analyze the evolution of the IMBH orbit after it kicked
out the neighboring stars. In order to do so, we performed another set
of simulations (run Bx), in which we placed the IMBH initially at the
distance 1/10 of that in runs Ax. Also, the stellar distribution is
cut off at smaller radius, to reduce the total number of particles. We
performed runs B1 through B4. The mass of the field stars in run B1 is
the same as that in A4. In B4, the mass ratio between the IMBH and field
stars is $10^3$, and therefore the mass of field stars is around the
solar mass. Two-body effects in run B4 is not larger than what would
occur in real galaxies.

\subsection{Evolution of the semi-major axis}
\label{sec:results-2nd-a}
\begin{figure}[htpb]
\plotone{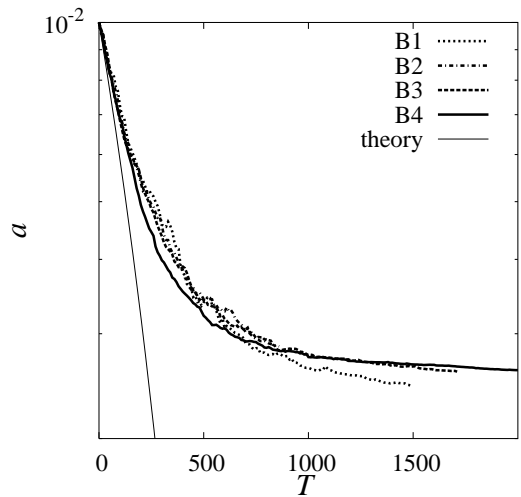}
\caption{Evolution of the semi-major axis of 
the IMBH. Dotted, dash-dotted, dashed, and solid 
curves are the results of models B1-4. Thin curve 
shows the theoretical curve calculated using
equation (\ref{eq:theoretical_a}). 
\label{fig:2nd-a}}
\end{figure}

Figure \ref{fig:2nd-a} shows the evolution of semi-major axis. We can
see that the evolution does not depend on the mass of field
stars. This result is not surprising since even for run B1 the
initial relaxation time of field particles at the initial location
of the IMBH is $7.3\times 10^4$ in the time unit 
and is much longer than
the duration of the calculation. The slowing down is much more
pronounced compared to that in runs Ax, simply because we followed the
evolution of the IMBH down to smaller value of $a$. The
evolution of the semi-major axis got stuck by $T=600$, when the
semi-major axis reached $a\sim 0.003$. 

\begin{figure}[htpb]
\plotone{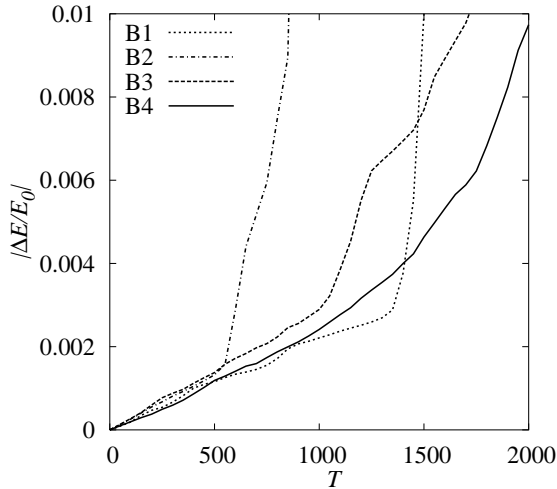}
\caption{Energy error of the calculations. 
The curves give the results for runs Bx. 
\label{fig:2nd-energy}}
\end{figure}

Figure \ref{fig:2nd-energy} shows the energy error for each of Bx
runs. Errors are reasonably small before $T=500$ for all runs. 
Quick increases in error are due to the increase in 
the eccentricity of the IMBH. 
At $T=500$, the binding energy of the IMBH is $\sim 60\%$ of the initial 
total binding energy of the system. 
Thus, relative energy error of $1\%$ corresponds to the
error of the semi-major axis of $1.6\%$, which is small compared to the
overall change of the semi-major axis.

\begin{figure}[htpb]
\plotone{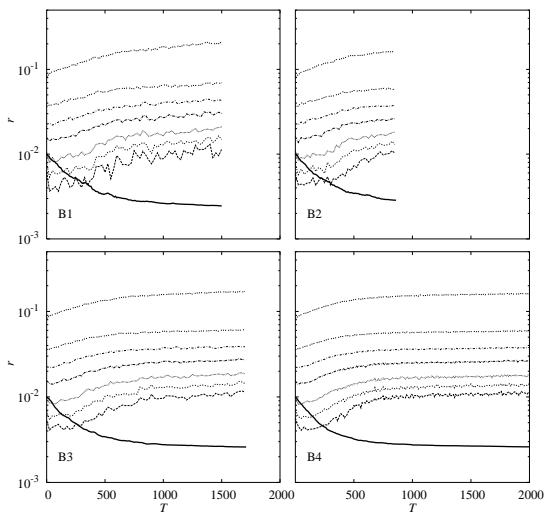}
\caption{Lagrangian radii around SMBH.
From bottom to top, the radii enclose 
$5\times10^{-5},$ 
$10^{-4}, 2\times10^{-4}, 5\times10^{-4},$
$10^{-3}, 2\times10^{-3}, 5\times10^{-3},$
0.01 and 0.02 of the unit mass. Thick solid lines 
indicate the semi-major axis of the IMBH. 
\label{fig:2nd-dense}}
\end{figure}

Figure \ref{fig:2nd-dense} shows the evolution of the Lagrangian
radii of field particles. As the IMBH sinks toward the SMBH, field 
stars expand. As the evolution of the IMBH slows down, the expansion 
of field stars also slows down. It is clear that the expansion, or 
depletion of the loss cone, is the cause of the slowing down of the 
IMBH orbit evolution.

\epsscale{0.5}
\begin{figure}[htpb]
\plotone{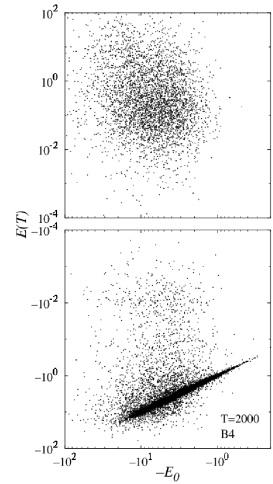}
\caption{Distribution of the field stars in 
the ($E_{0}, E(T)$) plane for model B3 at $T=2000$. 
Top and bottom panels show the particles with 
positive and negative energies, respectively.
\label{fig:2nd-ee}}
\end{figure}
\epsscale{1.0}

Figure \ref{fig:2nd-dense} might give the impression that all field
stars expand outward. Actually, that impression is wrong. 
Figure \ref{fig:2nd-ee} shows the initial and final energies
of field stars for run B3. Large fraction of field stars show very
small change in the energy, and some particles get very large
energy. This is of course the same as what is visible in figure 
\ref{fig:1st-ee} for run A4. Only the stars which strongly interacted 
with the IMBH are ejected, and other stars are essentially unaffected. 
The apparent expansion of outer Lagrangian radii in figure 
\ref{fig:2nd-dense} is the result of the ejection of the inner part 
of the field stars.

\subsection{Evolution of angular momentum}
\label{sec:results-2nd-L}

\begin{figure}[htpb]
\plotone{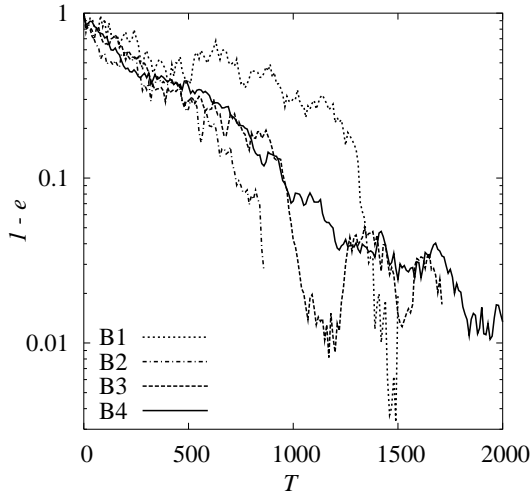}
\caption{Evolution of the eccentricity of the black 
hole binary. The curves give the results for runs 
Bx. We plot $(1-e)$ for the vertical axis.
\label{fig:2nd-e}}
\end{figure}

Figure \ref{fig:2nd-e} shows the evolution of eccentricities for runs
B1 to B4. Unlike the case of runs Ax, here it seems clear that in all
runs eccentricity shows slow increase in the early phase ($T<800$),
and in some runs there are phases when $e$ approaches to very close to
unity. However, exactly when such phase occurs shows large run-to-run
variation.

This quick increase of the eccentricity is rather surprising, since
in previous studies of massive BH binaries \citep{makino2004,
chatterjee2003} such increase has not been observed. However, since 
the configuration of the system is quite different, evolution can be 
very different. In the case of SMBH binary, two massive binaries have
similar mass, and field stars which interact with the binary are not
strongly bound to the binary. However, in our case of SMBH-IMBH
binary, all field stars are strongly bound to the SMBH, and have
almost Keplarian orbits. Thus, celestial-mechanical effects such as
the mean-motion resonance and Kozai mechanism \citep{kozai1962} can 
play important roles. These effects, on average, work as the 
transport mechanism for angular momentum from rapidly-rotating 
objects to slowly-rotating objects. In other words, these effects 
tend to increase the eccentricity of the IMBH. 

\begin{figure}[htpb]
\plotone{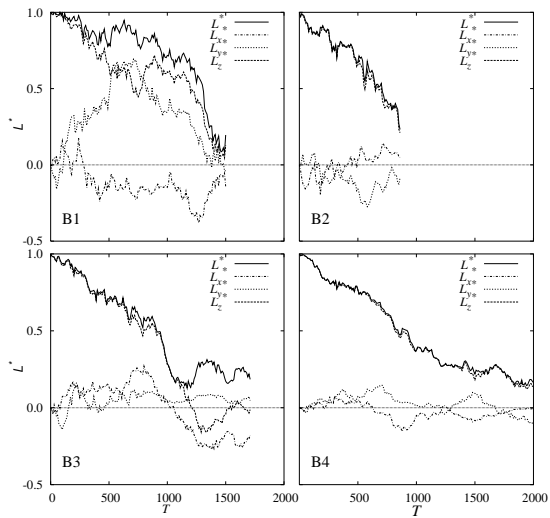}
\caption{Evolution of the normalized angular 
momentum of the BH binary. Dotted, dash-dotted, 
dashed, and solid curves give the results for 
$L_x^*, L_y^*, L_z^*$, and $L^*$, respectively.
\label{fig:2nd-L}}
\end{figure}

Figures \ref{fig:2nd-L} show the normalized angular momentum of the IMBH, 
$L^\ast \equiv L / L_0$, where $L$ is the angular momentum calculated in 
Equation (\ref{eq:ecc}) and $L_0 \equiv \sqrt{(M_S+M_I)a}$ is the
angular momentum of the circular orbit with the same semi-major axis. 
We can see that the slow increase of the eccentricity directly
corresponds to the decrease of the total angular momentum, and the
random walk of the angular momentum vector is small, especially for
runs with large $N$. Apparently, only after the quick increase of the
eccentricity the random walk of the angular momentum vector becomes
noticeable. Thus, probably the high eccentricity state corresponds to
some statistical equilibrium.

\subsection{Why the eccentricity goes up?}
Here we try to understand the mechanism which drives the increase of
the eccentricity. We concentrate on run B4, since it has the largest
number of field particles and therefore the statistical analysis of
the behavior of field particles is the most reliable. In particular,
as we can see in figure \ref{fig:2nd-L}, in run B4 the orbital plane
of the IMBH remains close to the xy-plane, while in other runs the
random walk of the angular momentum vector is significant.

In order to understand the behavior of the IMBH, it would be useful
to see with which field particles the IMBH interacted. Figure 
\ref{fig:a-cdl-1} shows the cumulative change of angular momentum of
field particles as the function of their semi-major axis $a$. Here,
the change in the angular momentum $\Delta l$ of a particle is 
defined as
\begin{equation}
\Delta l \equiv [{\bf L}(T) - {\bf L}(T +
\Delta T)]\cdot \hat{\bf L}_{\rm IMBH}~, \label{eq:dl_0}
\end{equation}
where ${\bf L}(T)$ is its angular momentum vector at time $T$, and
$\hat{\bf L}_{\rm IMBH}$ is the unit vector with the direction of the
angular momentum vector of the IMBH. We use $\Delta T = 50$ for all
values of $T$. Thus this $\Delta l$ is the angular momentum
change projected to the orbital plane of the IMBH.

\begin{figure}[htpb]
\plotone{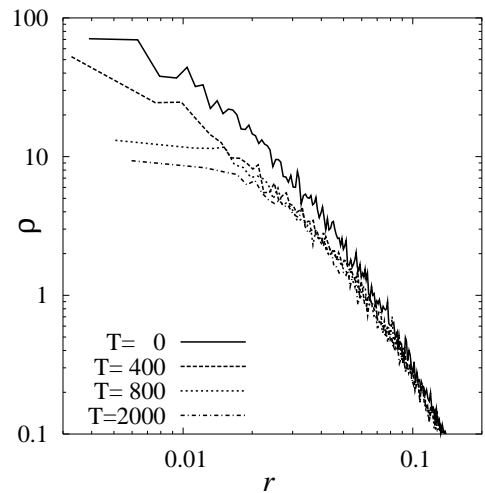}
\caption{Radial density profiles at times 
$T=0$, 400, 800, and $2000$ for run B4. 
\label{fig:rho2}}
\end{figure}

Figure \ref{fig:rho2} show the change of the density profiles for 
run B4. We can see that the central density decreases, 
and the inclination of the central cusp becomes flat. 
It is caused by the ejection of field stars by the IMBH.

\begin{figure}[htpb]
\plotone{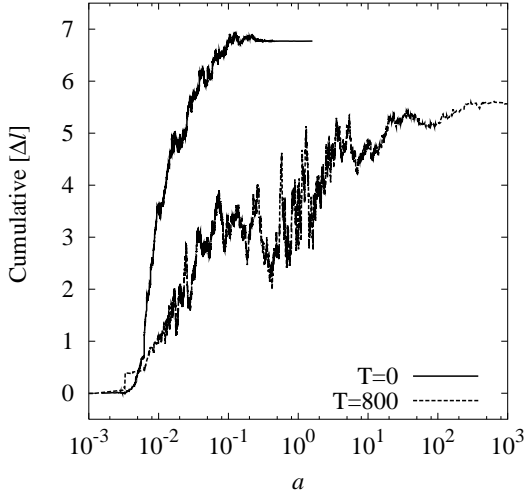}
\caption{Cumulative change of angular momentum of 
field particles $\Delta l$ as a function of 
semi-major axis $a$ for the times $T=0$ and $800$.
\label{fig:a-cdl-1}}
\end{figure}

The difference between the plot for $T=0$ and $T=800$ is that, even
though the semi-major axis of the IMBH is much smaller at $T=800$,
the field particles which exchanged the angular momentum has much
wider distribution at $T=800$ than at $T=0$. For $T=0$, stars with the
semi-major axis comparable to that of the IMBH dominate the change in the
angular momentum, while for $T=800$, stars with $a$ comparable to that
of the IMBH (0.003) have practically no contribution, and effect of stars
with $a>1$ is not negligible. Of course, since the IMBH ejected most
of stars with $a<0.01$, it cannot interact with them. 

\begin{figure}[htpb]
\plotone{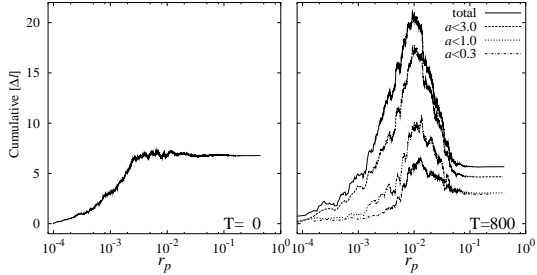}
\caption{Same as figure \protect \ref{fig:a-cdl-1} 
but as a function of the periastron distance $r_p$ 
for the times $T=0$ and $800$. In the right hand 
panel, curves are, from top to bottom, total change, 
changes for particles with $a<$3.0, 1.0, and 0.3.
\label{fig:r-cdl-1}}
\end{figure}

Figure \ref{fig:r-cdl-1} shows the same cumulative plot but as the
function of the periastron distance $r_p$. For $T=0$, particles which
go inside the IMBH orbit account for the change in the angular
momentum. On the other hand, at $T=800$ particles orbit outside the
IMBH orbit are responsible for the change in the angular momentum, and
that tendency is more significant for field particles with small $a$.

For field stars with relatively small $a$, we can draw the following
picture. The IMBH has created the hole in the distribution function, 
and it can interact only with stars with semi-major axis larger or
comparable to its apocenter distance. In other words, the IMBH can 
strongly interact with other stars only when it is at the apocenter. 
If it loses kinetic energy at its apocenter, it becomes strongly
eccentric.

\begin{figure}[htpb]
\plottwo{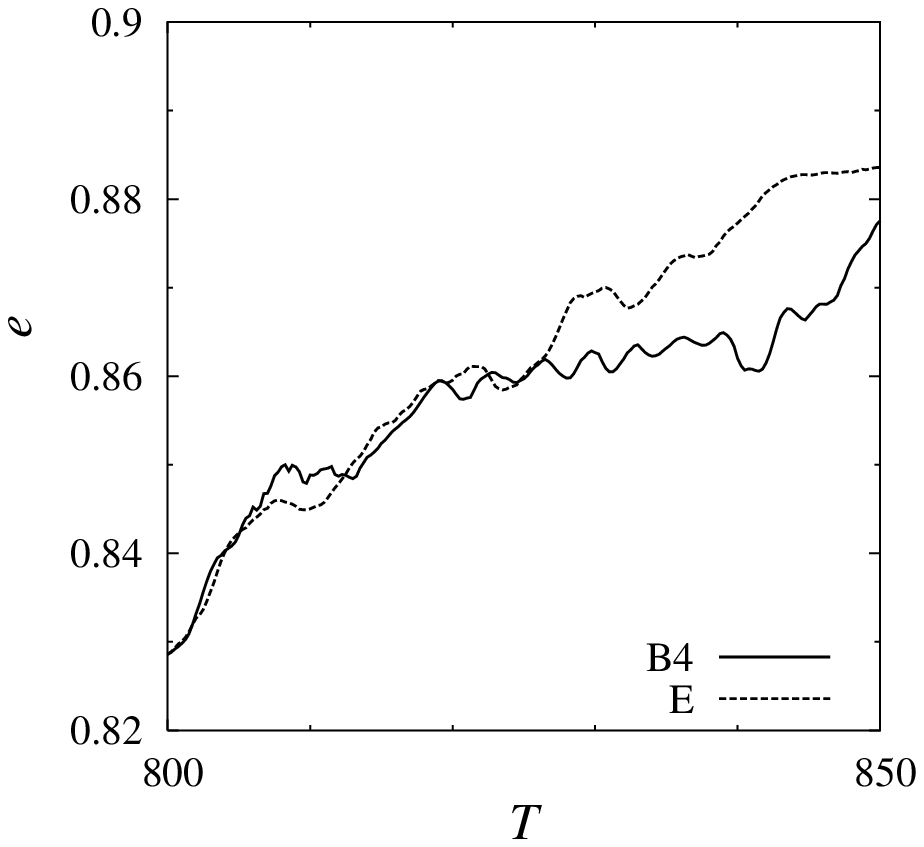}{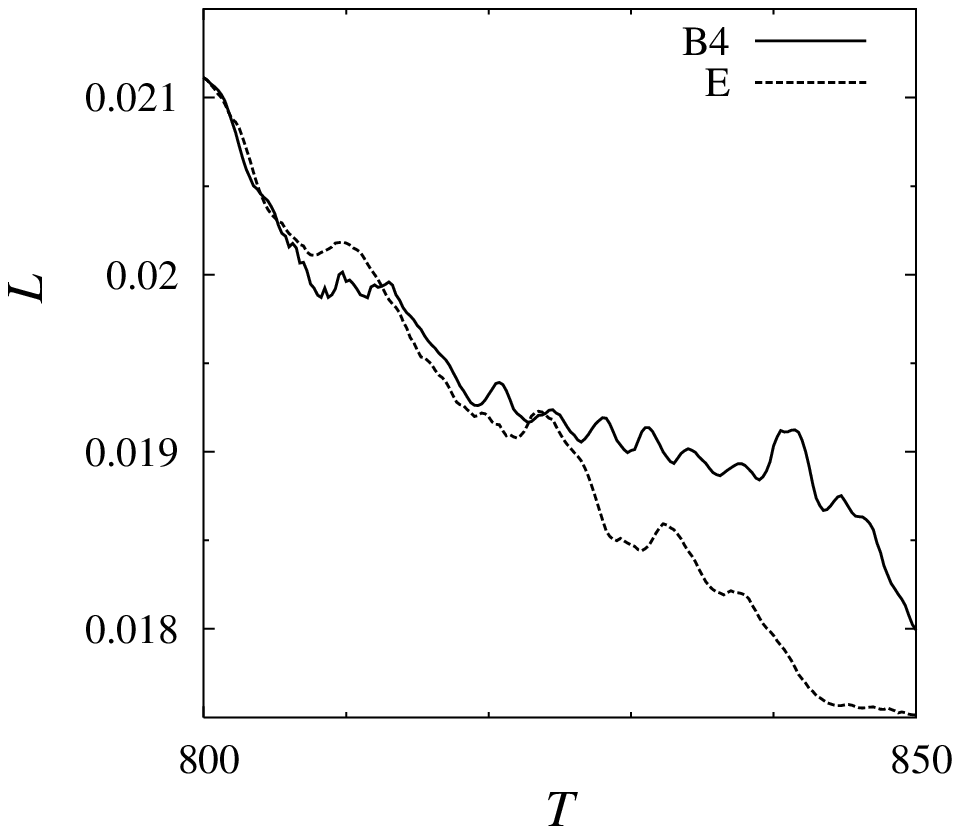}
\caption{Comparison of run B4 and run E. As run E, 
we started from the $T=800$ snapshot of run B4 but 
we removed all stars with $a>0.3$. Left is 
the evolution of the eccentricity, and right is 
the evolution of the angular momentum of the BH binary. 
\label{fig:run-e}}
\end{figure}

Interaction with stars with large $a$ is more complex. However, here
the interaction can occur at positions other than the apocenter, since
field particles with large $a$ can have small $r_p$. Thus, the
increase of the eccentricity is likely to be driven by interactions
with field particles with small $a$. In order to test this hypothesis,
we performed one additional run, run E, which we started from the
$T=800$ snapshot of run B4 but we removed all stars with $a>0.3$. 
Fig \ref{fig:run-e} shows the result. The change in the eccentricity
is largely similar. So we can conclude that the interactions with field
stars with small $a$ (and relatively large $r_p$) drive the
increase of the eccentricity of the IMBH. 

This mechanism of the increase of the eccentricity is similar to that 
proposed by \citet{fukushige1992}. They argued that the dynamical
friction on an eccentric binary should be most effective at the
apocenter and therefore an eccentric binary should become more
eccentric. For the binary SMBH of comparable masses, this mechanism
turned out to be ineffective, because the change in the orbital
elements of a binary is really the result of rather complex three-body
interaction between two binary components and the incoming
star. However, in our case of IMBH-SMBH binary, the situation is very
different. Since the mass of the IMBH is much smaller than that of SMBH,
field stars must come close to the IMBH to change its orbit. Thus, the
IMBH
does have more chance to interact with field starts at apocenter than
at pericenter, and its eccentricity increases.

\section{Discussion and Summary}

\subsection{Final fate of IMBH}
\label{sec:gravitational-radiation}
We have seen that the evolution of semi-major axis of the IMBH effectively
stops, when it ejected field stars in that region. We have also seen
that the eccentricity of the IMBH, after the evolution of semi-major axis
stopped, can reach very close to unity. From the viewpoint of the evolution 
of SMBH, important question is what the final fate of the IMBH is.

\begin{figure}[htpb]
\plotone{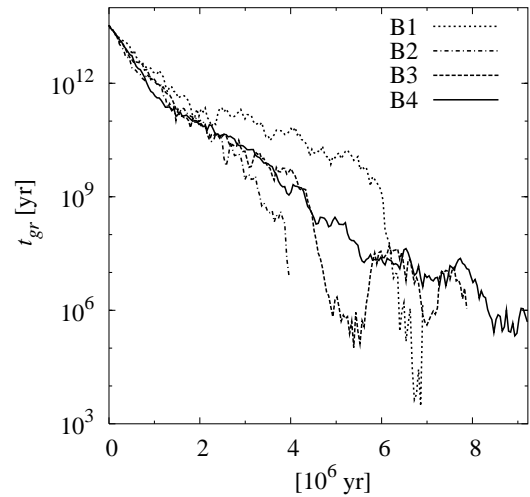}
\caption{Gravitational radiation timescale 
given by equation (\ref{eq:t-gr2}) for runs Bx.
\label{fig:2nd-tgr}}
\end{figure}

Figure \ref{fig:2nd-tgr} shows the timescale of the merging through 
gravitational wave radiation, for runs Bx. We converted the simulation 
units to physical units.

The timescale of the merging through gravitational wave radiation is given by
\begin{eqnarray}
t_{\rm gr} &=& {5 \over 256} 
\frac{a^{4} c^{5}}{ G^{3}M_{S}M_{I}(M_S+M_I)} F(e)
\label{eq:t-gr}
\end{eqnarray}
with $F(e)$ is
\begin{eqnarray}
F(e) &=& \left( {1-e^{2}}\right)^{7/2}
\left( { 1+ \frac{73}{24} e^{2} + \frac{37}{96}e^{4}}\right)^{-1} 
\label{eq:fe}
\end{eqnarray}
\citep{peters1964}. 
The timescale  $t_{\rm gr}$ depends strongly on $e$, when $e$ is
close to unity. In our units, $t_{\rm gr}$ is given by 
\begin{eqnarray}
t_{gr} &\simeq & 6.3\times 10^{13} F(e)\nonumber \\
&&
\left( {a \over 0.01 \rm pc}\right)^4
\left( {M_S \over 3\times\! 10^{6} M_{\odot}}\right)^{-2} \nonumber \\
&&
\left( {M_I \over 3\times\! 10^{3} M_{\odot}}\right)^{-1} {\rm yr}. 
\label{eq:t-gr2}
\end{eqnarray}

We can see from figure \ref{fig:2nd-tgr} that the merging timescale
can go down to less than $10^6$ yrs, and stays at that value for some
time. In other words, if we take into account the effect of
gravitational wave radiation, the IMBH will merge to SMBH.

\subsection{Event rate for gravitational wave detection}

\citet{matsubayashi2004} estimated the event rate of the merging of
IMBH-SMBH binary, assuming that all binaries eventually merge. Our
present result indicates that this assumption of 100\% merging is
valid.  They estimated the event rate to be $20 \sim 70$ per year, for
the detection limit of $h \approx 10^{-21}$. Here, $h$ is the
dimensionless amplitude of gravitational wave.
The frequency of the gravitational wave in their final merging phase
is $10^{-1}$ to $10^{2}$ Hz. It is within the target range of
LISA\citep{LISA} and DECIGO\citep{seto01}.

\citet{matsubayashi2004} considered two limiting cases for the growth
of SMBH. In the first case, the growth is hierarchical. BHs always
merge with another BH with a similar mass. In the second case, the
growth is monopolistic and one BH always merge with IMBH of small
mass. In the first case, large number of events has rather low
amplitude, since they come from IMBH-IMBH mergings. In the second
case, most events come from IMBH-SMBH merging, and therefore amplitude
is bigger than that in the first case. DECIGO will be able to detect
all events for both cases, while LISA would not detect the majority of
events in the case of hierarchical growth. In the case of 
the monopolistic growth, both DECIGO and LISA will detect most of events.
Thus, DECIGO event rate would be around 50, while LISA event rate
would be 5-50, depending on the growth mode.

\subsection{Summary}

In this paper, we studied the orbital evolution of the IMBH after its
parent cluster is completely disrupted. Our main findings are
summarized as follows.

Initially, the IMBH sink toward the SMBH through dynamical friction. 
However, the evolution of the semi-major axis stops when the IMBH 
approaches to the radius at which the initial stellar mass is comparable 
to the IMBH mass. For our galaxy, it corresponds to around 0.01 pc. If 
the IMBH remains in a circular orbit, the merging timescale by 
gravitational wave radiation would be $6.3\times 10^{13}$ yrs. 

However, in this region the eccentricity of the IMBH approaches to unity, 
and therefore we expect the IMBH to quickly merge with the SMBH. This 
increase is due to interactions with field stars with periastron larger 
than the semi-major axis of the IMBH. The fact that many of the stars 
very close to the galactic center have large eccentricities is probably 
explained by the same mechanism. For example, S2 has $e\sim 0.87$ and 
$a\sim 0.119'' (\simeq 4\times 10^{-3} pc)$\citep{schodel2003}. 
The distribution of eccentricities of known stars very close to SgrA* 
is not consistent with being isotropic.

\acknowledgments
We thanks Masaki Iwasawa and Keigo Nitadori for simulating discussions. 
We are also grateful to Shigeru Katagiri for preparing the research 
environment. 

This research is partially supported by the Special Coordination Fund
for Promoting Science and Technology (GRAPE-DR project), Ministry of
Education, Culture, Sports, Science and Technology, Japan.

{}


\begin{thebibliography}{}

\bibitem[Alcubierre {\it et al.}(2001)]{alcubierre2001}
Alcubierre, M., Bernger, W., Bruegmann, B., Lanfermann,  G.,
Nerger, L., Seidel, E., and Takahashi, R., 2001, \prl, 87, 271103
\bibitem[Bahcall \& Wolf (1976)]{bahcall1976}
Bahcall, J. N., \& Wolf. R. A., 1976, \apj, 209, 214
\bibitem[Binney \& Tremaine(1987)]{BT}
Binney, J. J., \& Tremaine, S., 1987, Galactic Dynamics (Princeton University Princeton Press)
\bibitem[Begleman, Blandford, \& Rees(1980)]{BBR1980}
Begelman, M. C., Blandford, R. D., \& Rees, M. J., 1980, \nat, 287, 307
\bibitem[Berczik {\it et al.}(2005)]{berczik2005}
Berczik,P., Merritt, D., \& Spurzem, R., 2005, \apj 633, 680
\bibitem[Ebisuzaki {\it et al.}(2001)]{ebisuzaki2001}
Ebisuzaki, T., Makino, J., Tsuru, T.G., Funato, Y, Zwart, S.P., Hut, P.,
McMillan, S., Matsushita, S., Matsumoto, H., \& Kawabe, R. 2001, \apjl, 562, L19
\bibitem[Fukushige {\it et al.}(2005)]{fukushige2005}
Fukushige, T., Makino, J., \& Kawai, A., 2005, \pasj, in press, astro-ph/0504407
\bibitem[Fukushige {\it et al.}(1992)]{fukushige1992}
Fukushige, T., Ebisuzaki, \& T., Makino, J., 1992, \pasj, 44, 281
\bibitem[kozai (1962)]{kozai1962}
Kozai, Y., 1962, \aj, 67, 591
\bibitem[Chatterjee {\it et al.}(2003)]{chatterjee2003}
Chatterjee, P., Hernquist, L., \& Loeb, A., 2003, \apj, 592, 32
\bibitem[Guesten {\it et al.}(1987)]{guesten1987}
Guesten, R., Genzel, R., Wright, M. C., Jaffe, D. T., Stuzki, J., \& Harris, A. I., 1987, \apj, 318, 124
\bibitem[Genzel {\it et al.}(1996)]{genzel1996}
Genzel, R., Thatte, N., Krabbe, A., Kroker, H., \& Tacconi-Garman, L. E., 1996, \apj, 472, 153
\bibitem[Genzel {\it et al.}(1997)]{genzel1997}
Genzel, R., Eckart, A., Ott, T., \& Eisenhauer, F., 1997, \mnras, 291, 219
\bibitem[Genzel {\it et al.}(2000)]{genzel2000}
Genzel, R., Pichon, C., Eckart, A., Gerhard, O. E., \& Ott, T., 2000, \mnras, 317, 348
\bibitem[Hashimoto {\it et al.}(2003)]{hashimoto2003}
Hashimoto, Y., Funato, Y., \& Makino, J., 2003, \apj, 582, 196
\bibitem[Kawakatu (2002)]{kawakatu2002}
Kawakatu, N., \& Umemura, M., 2002, \mnras, 329, 572
\bibitem[LISA report(2000)]{LISA}
LISA Laser Interferometer Space Antenna: A Cornerstone Mission for the Observation of Gravitational Waves, System \& Technology Study Report, ESA-SCI (2000) 11
\bibitem[Makino \& Aarseth (1992)]{makino1992}
Makino, J., \& Aarseth, S. J., 1992, \pasj, 44, 141
\bibitem[Makino (1997)]{makino1997}
Makino, J., 1997., \apj, 478, 58 
\bibitem[Makino {\it et al.}(2003)]{makino2003}
Makino, J., Fukushige, T., Koga, M., \& Namura, K., 2003, \pasj, 55, 1163
\bibitem[Makino \& Funato (2004)]{makino2004}
Makino, J., \& Funato, Y. 2004, \apj, 602, 93
\bibitem[Matsubayashi {\it et al.}(2004)]{matsubayashi2004} 
Matsubayashi, T., Shinkai, H., \& Ebisuzaki, T., 2004, \apj, 614, 864 
\bibitem[Matsumoto {\it et al.}(2001)]{matsumoto2001}
Matsumoto, H., Tsuru, T. G., Koyama, K., Awaki, H., Canizares, C. R., Kawai, N., Matsushita, S., \& Kawabe, R., 2001, \apjl, 547, L25
\bibitem[Matsushita {\it et al.}(2000)]{matsushita2000}
Matsushita, S., Kawabe, R., Matsumoto, H., Tsuru, T. G., Kohno, K., Morita, K., Okumura, S. K., \& Vila-Vilaro, B., 2000, \apjl, 545, L107
\bibitem[Maillard {\it et al.}(2004)]{maillard2004}
Maillard, J. P., Paumard, T., Stolovy, S. R., \& Rigaut, F., 2004, A\&A, 423, 155
\bibitem[McCrady {\it et al.}(2003)]{mccrady2003}
McCrady, N., Gilbert, A. M., \& Graham, J. R., 2003, \apj, 596, 240
\bibitem[McLure \& Dunlop(2001)]{mclure2001}
McLure, R. J., \& Dunlop, J. S., 2001, \mnras, 321, 515
\bibitem[Miller \& Hamilton (2002)]{miller2002}
Miller, M. C., \& Hamilton, D. P., 2002, \mnras, 330, 232
\bibitem[Milosavljevi${\rm\acute{c}}$ \& Merritt (2001)]{milos2001}
Milosavljevi${\rm\acute{c}}$, M., \& Merritt, D., 2001,\apj, 563, 34
\bibitem[Milosavljevi${\rm\acute{c}}$ \& Merritt (2003)]{milos2003}
Milosavljevi${\rm\acute{c}}$, M., \& Merritt, D., 2003,\apj, 596, 860
\bibitem[Peters (1964)]{peters1964}
Peters, P. C., 1964, \prb, 136, 1224
\bibitem[Portegies Zwart {\it et al.}(2002)]{simon2002}
Portegies Zwart, S. F., Makino, J., McMillan, S. L. W., \& Hut, P., 2002, \apj, 565, 265
\bibitem[Portegies Zwart {\it et al.}(2004)]{simon2004}
Portegies Zwart, S. F., Holger, B., Makino, J., McMillan, S. L. W., \& Hut, P., 2004, \nat, 428, 724
\bibitem[Portegies Zwart {\it et al.}(2005)]{simon2005}
Portegies Zwart, S. F., Holger, B., McMillan, S. L. W., Makino, J., Hut, P., \& Ebisuzaki, T., 2005, astro-ph/0511397 
\bibitem[Rees(1990)]{Rees1990}
Rees, M. J., 1990, {\it Science}, 247, 817
\bibitem[Seto {\it et al.}(2001)]{seto01}
Seto, S., Nakamura, T., \& Kawamura, S., 2001, \prl, 87, 221103
\bibitem[Szell \& Merritt (2005)]{szell2005}
Szell, A., \& Merritt, D., 2005, astro-ph/0502198
\bibitem[Schodel {\it et al.}(2003)]{schodel2003} 
Schodel, R., Ott, T., Genzel, R., Eckart, A., Mouawad, N.,\& Alexander, T., 2003, \nat, 596, 1015
\bibitem[Taniguchi {\it et al.}(2000)]{taniguchi2000} 
Taniguchi, Y., Shioya, Y., Tsuru, T. G., \& Ikeuchi, S., 2000, \pasj , 52, 533
\bibitem[Tremaine {\it et al.}(1994)]{tremaine1994}
Tremaine, S., et al., 1994, \aj, 107, 634
\bibitem[Quinlan \& Hernquist(1997)]{quinlan1997}
Quinlan, G. D., \& Hernquist, L.,1997, {\it New Astronomy}, 2, 533
\end{thebibliography}
\end{document}